\documentclass[preprint,review,12pt]{elsarticle}

\AtBeginDocument{%
	\providecommand\BibTeX{{%
			\normalfont B\kern-0.5em{\scshape i\kern-0.25em b}\kern-0.8em\TeX}}}

\usepackage{geometry}
\usepackage{subcaption}
\usepackage[mathscr]{eucal} 
\usepackage{amsmath} 
\usepackage{algorithmic} 
\usepackage[linesnumbered,lined,boxed,commentsnumbered,ruled]{algorithm2e}
\usepackage{booktabs}
\usepackage{array,graphicx, multirow} 
\newcommand{\STAB}[1]{\begin{tabular}{@{}c@{}}#1\end{tabular}}

\newcommand{\setItemSep}{\setlength\itemsep}

\usepackage[table,xcdraw, prologue]{xcolor}
\usepackage{nicefrac}
\usepackage{optidef}
\usepackage{tikz}
\usetikzlibrary{matrix}
\usepackage[shortlabels]{enumitem}
\RequirePackage[pdftex,
colorlinks={true},
linkcolor={Blue3},
urlcolor={Blue3},
citecolor={Blue3},
pdfstartview={FitH},
bookmarks={true},
pdfauthor={Author},
pdftitle={Title},
pdfsubject={Subject}
plainpages=false,
pdfpagelabels
]{hyperref}
\journal{EASE'21}

\begin{document}
	
	\title{DABT: A Dependency-aware Bug Triaging Method}
	
	\author[1]{Hadi Jahanshahi}
	\ead{hadi.jahanshahi@ryerson.ca}
	\author[2]{Kritika Chhabra}
	\author[1]{Mucahit Cevik}
	\author[1]{Ay\c{s}e Ba\c{s}ar}
	
	\address[1]{Data Science Lab at Ryerson University, Toronto, Canada}
	
	\address[2]{Data Science and Analytics, Ryerson University, Toronto, Canada}

	\begin{abstract}
		In software engineering practice, fixing a bug promptly reduces the associated costs. 
		On the other hand, the manual bug fixing process can be time-consuming, cumbersome, and error-prone. 
		In this work, we introduce a bug triaging method, called Dependency-aware Bug Triaging (DABT), which leverages natural language processing and integer programming to assign bugs to appropriate developers. 
		Unlike previous works that mainly focus on one aspect of the bug reports, DABT considers the textual information, cost associated with each bug, and dependency among them.
		Therefore, this comprehensive formulation covers the most important aspect of the previous works while considering the blocking effect of the bugs. 
		We report the performance of the algorithm on three open-source software systems, i.e., {\scshape{EclipseJDT}}, {\scshape{LibreOffice}}, and {\scshape{Mozilla}}. 
		Our result shows that DABT is able to reduce the number of overdue bugs up to 12\%. 
		It also decreases the average fixing time of the bugs by half. 
		Moreover, it reduces the complexity of the bug dependency graph by prioritizing blocking bugs. 
		
	\end{abstract}
	

	\maketitle
	

	\section{Introduction}
	In software engineering practice, fixing a bug promptly reduces the associated costs~\citep{Kumar2017}. 
	The manual bug fixing process can be time-consuming, cumbersome, and error-prone. 
	Therefore, many researchers investigate the possibility of automating each component of this process. 
	When a bug is reported to an issue tracking system (ITS), triagers start exploring its validity and, accordingly, decide on how to proceed. 
	If a bug report does not include enough information for reproducibility, or if it is not relevant, it will be flagged as an invalid bug. 
	However, for valid bugs, triagers may find a proper developer to assign the existing bug. 
	In many open software systems, developers themselves may claim a bug's possession and start working on it. 
	These decisions are mainly made by checking the bug's information, e.g., its title, component, description and severity. 
	
	Previous studies mainly concentrate on the importance of textual information in the bug triaging process~\cite{mani2019deeptriage, anvik2006should, lee2017applying}, . 
	Considering the assignment task as a classical classification problem, some researchers explored the effect of considering different bug report features -- e.g., title, description, and tag -- on assigning a proper developer.
	However, such approaches fail to address the issue of longer fixing times of some bugs that are blindly assigned. 
	Other works also considered the combination of bug-fixing time (cost) and selecting the proper developer (accuracy)~\cite{park2011costriage, kashiwa2020}.
	They use a regulatory term/parameter and suggest that their approach reduces the fixing time without compromising the accuracy. 
	However, the algorithms still suffer from task concentration, i.e., assigning an unmanageable number of bugs to the highly expert developers. 
	A recent study by~\citet{kashiwa2020} proposes setting an upper limit on the number of tasks that each developer can address given a predefined period of time. 
	Their method alleviated the problem of imbalanced bug distribution and reduced the number of overdue bugs.
	However, in their objective function, they do not consider the bug fixing time. 
	Also, their model, similar to previous works, does not consider the significant effect of blocking bugs in the prioritization task~\citep{akbarinasaji2018partially}. 
	We propose a Dependency-aware Bug Triaging method (DABT) that considers both the bug dependency and the fixing time during the bug triage process. 
	Accordingly, DABT employs machine learning algorithms and integer programming to determine the suitable developers given their available time slots and bug fixing times. 
	
	We organized the rest of the paper as follows. Section~\ref{sec:background} briefly discusses the background of relevant methods used on our proposed bug triage algorithm, which is followed by a motivating example in Section~\ref{sec:motivating_example}.
	Section~\ref{sec:research-methodology} describes the DABT methodology and bug report dataset used in our work. Section~\ref{sec:results} investigates the research questions while taking into account the importance of the bug dependency. Finally, Section \ref{sec:threats} describes the limitations and threats to validity, and Section \ref{sec:conclusion} concludes the paper.

	\section{Background}\label{sec:background}
	In this section, we briefly discuss the models employed in our proposed approach. Specifically, we adopt Latent Dirichlet allocation (LDA) for topic modeling over bug descriptions and Integer Programming (IP) for bug triaging. 
	Note that, for the developer assignment task, we use Support Vector Machines (SVM), whose details were skipped as it is a widely used model in the domain.
	
	\subsection{LDA}
	LDA~\citet{blei2003lDA} is a probability-based topic modeling method. 
	It is an unsupervised learning approach that can identify the topics in given documents or corpus based on their word clusters and frequencies. 
	LDA assumes that the document is a mixture of topics, and each word is attributed to one of the topics. 
	In the process of bug triaging, LDA is typically used to identify the bug type given a bug report, where the latter refers to documents, and the former refers to a topic. 
	That is, we extract bug description and summary from a bug report and create a bag of words (BOW) after removing the stop words. 
	Then, given the BOW of each document, LDA arbitrarily assigns each word of the report to one of the topics (or bug category). 
	Next, it iteratively reassigns each word, assuming that other words are accurately allocated. 
	Accordingly, it computes the two conditional probabilities, the proportion of words in document $d$ that are assigned to topic $t$--i.e., $p(\text{topic}_t|\text{document}_d)$-- and proportion of assignments to topic $t$ over all documents that come from this word as $p(\text{word}_w| \text{topic}_t)$. 
	The product of these conditional probabilities gives a probability distribution based on which a new topic is assigned.
	
	\subsection{IP modeling}
	Integer programming is a mathematical modeling framework for optimization problems where certain decision variables need to be integer-valued.
	It has a wide range of applications in various domains including healthcare, energy and manufacturing \citep{chen2010applied}.
	Thanks to major advances in integer programming solution methodologies and their integration into commercial and noncommercial solvers, they have become increasingly popular over time.
	It is particularly suited for scheduling problems which typically require various integer and binary decisions and a long list of constraints to be satisfied. For instance, \citet{sung2016optimal} formulate an IP model for the problem of resource-constrained triage in a mass casualty incident, where the priority of the patients is identified for deploying limited emergency medical service resources so that the maximum number of patients benefit from the response efforts. 
	
	IP modeling can be similarly employed for the bug triaging problem, which involves identifying the priority of the bugs and the most suitable developers to assign those. That is, through an appropriate mathematical model, various problem constraints (e.g., blocking bugs and deadlines) can be handled while optimizing over a particular objective, such as maximizing the match between the bug descriptions and the developers. This way, the bug triaging can be automated to a certain extent while the overall efficiency of the process is enhanced.
	

	\subsection{Overview of the existing methods}\label{sec:existing-methods}
	In a typical bug tracking system, after a bug is validated, the first step is to assign it to an appropriate developer. 
	Accordingly, the bug assignment is usually considered a critical task which prompted many researchers to work towards its automation. 
	The proposed methods for this problem can be categorized as content-based recommendation (CBR), cost-aware bug triaging (CosTriage), and release-aware bug triaging (RABT). 
	We compare our proposed approach with the representative methods from these categories.
	
	\subsubsection{Content-based recommendation}
	CBR approach aims to assign the most appropriate developer to the incoming bug through analyzing its content. 
	\citet{anvik2006should} used machine learning techniques to build a semi-automated bug triager. 
	They trained a multi-class classifier on the bug history by using SVM, Naive Bayes, and C4.5 trees, where bug title and description were converted into a feature vector as the input data, and the assigned developers were taken as the labels. 
	The resulting classifier then analyzes the textual contents of a given report and estimates each developer’s score for the new bug by extracting its feature vector and applying the classifier. 
	Therefore, it can recommend suitable developers for any newly-reported bugs. 
	As they reported SVM to show the best performance, we use it as the classifier in our study. 
	Nonetheless, some CBR studies followed the same concepts by training deep learning algorithms~\citep{lee2017applying, mani2019deeptriage, Zaidi2020}. 
	As these studies revolve around the same idea while reporting modest accuracy improvements, we consider the most commonly used approach in our analysis.

	\subsubsection{Cost-aware recommendation}
	CosTriage considers both the accuracy and the cost of an assignment.
	\citet{park2011costriage} proposed a bug-triage method combining an existing CBR with a collaborative filtering recommender (CF). 
	They model the developer profile that denotes their estimated cost for fixing a particular bug type. 
	To create these developer profiles, they quantified their profile values as the average bug fixing time per bug type, where bug types are determined by applying the LDA to bug summary and description. 
	Next, they estimated the suitability of each developer using CBR. 
	They trained SVM on the textual information of the bugs and tested it for new bugs in the system. 
	Finally, by combining the cost and the possibilities predicted by CBR, a bug is assigned to a developer with the highest score.
	
	\subsubsection{Release-aware recommendation}
	\citet{kashiwa2020} proposed the RABT method that primarily focuses on increasing the number of bugs resolved by the following release. 
	RABT considers the time available before the next release and simultaneously keeps track of the bug fixing load on a developer. 
	They formulated bug triage as a multiple knapsack problem to optimize the assignment task. 
	In the standard knapsack problem, for a given set of items, each with a weight and a value, the objective is to determine the number of each item to include in a collection (i.e., knapsack) so that the total weight is less than or equal to a given limit (i.e., knapsack capacity), and the total value is as large as possible.
	Similarly, the problem of assigning bugs to the developers can be considered as a variant of the knapsack problem.
	That is, RABT pairs the best possible combination of bugs and developers to maximize the bug-fixing efficiency given a time limit. 
	They used SVM and LDA to compute the preferences and costs for each bug and developer, respectively. 
	LDA, similar to CosTriage, categorizes the bug and calculates the average time taken for each developer to fix bugs in each category. 
	For a new bug, SVM computes developers' preferences, whereas LDA calculates their expected fixing time. 
	Then, RABT determines the available time slot $T_t^d$ for developer $d$ based on their bugs at hand and the project horizon $\mathcal{L}$ (predetermined for each developer).
	Ultimately, bugs are assigned to the most suitable developers such that their preferences are maximized, while their fixing cost does not exceed their available time slots.
	
	\subsubsection{Research gaps}
	The existing methods discussed above have certain limitations. 
	While CBR~\cite{anvik2006should} is a highly accurate approach, it is prone to over-specialization, recommending only bugs similar to what a developer has solved before. 
	Thus, it concentrates the task on some experienced developers and slowing down the bug fixing process. 
	In addition, it only considers the accuracy as the performance metric, ignoring all other parameters such as bug-fixing time, developer’s interest, and expertise. 
	The Costriage \cite{park2011costriage} method addresses the drawbacks of CBR and tries to optimize both the accuracy and bug fixing cost. 
	It estimates the bug fixing time using the LDA and overcomes its sparseness through a hybrid approach and collaborative filtering recommender. 
	It improves the bug fixing time without substantial degradation of accuracy.
	However, both methods disregard the number of bugs a developer can fix in a given time frame. 
	They may assign more bugs to experienced developers than they can address in the available time. 
	Moreover, they do not consider the number of bugs that a developer can fix before the next release. 
	
	\citet{kashiwa2020} focus on increasing the number of fixed bugs before the next release by setting an upper limit on the available time for each developer. 
	Accordingly, RABT mitigates the task concentration, assigns a more achievable number of bugs that a developer can fix in a given time, and reduces the bug fixing time. 
	Also, the order of assigned bugs impacts the number of bugs fixed by the release. 
	If the model initially triages a time-consuming bug, it can decrease the number of bugs fixed by the next release. 
	RABT also prioritizes the developers with fewer tasks as they are available to handle new bugs. 
	However, it reduces the accuracy of bug assignments. 
	Besides, setting the proper upper limit can be challenging. 
	If the model wants to focus on the number of bugs fixed before the next release, it should determine a dynamic threshold on developers' available times. 
	Practically, by setting a constant value, the model does not become linked to the release dates. 
	
	Another significant consideration ignored by all these models is the bug dependency and the impact of the blocking bugs. 
	That is, none of these models take into account bug dependency information during the triage process, which potentially leads to many infeasible tasks imposed on developers. 
	
	\section{Motivating example}\label{sec:motivating_example}
\begin{figure}[!ht]
		\centering
		\includegraphics[width=0.7\linewidth]{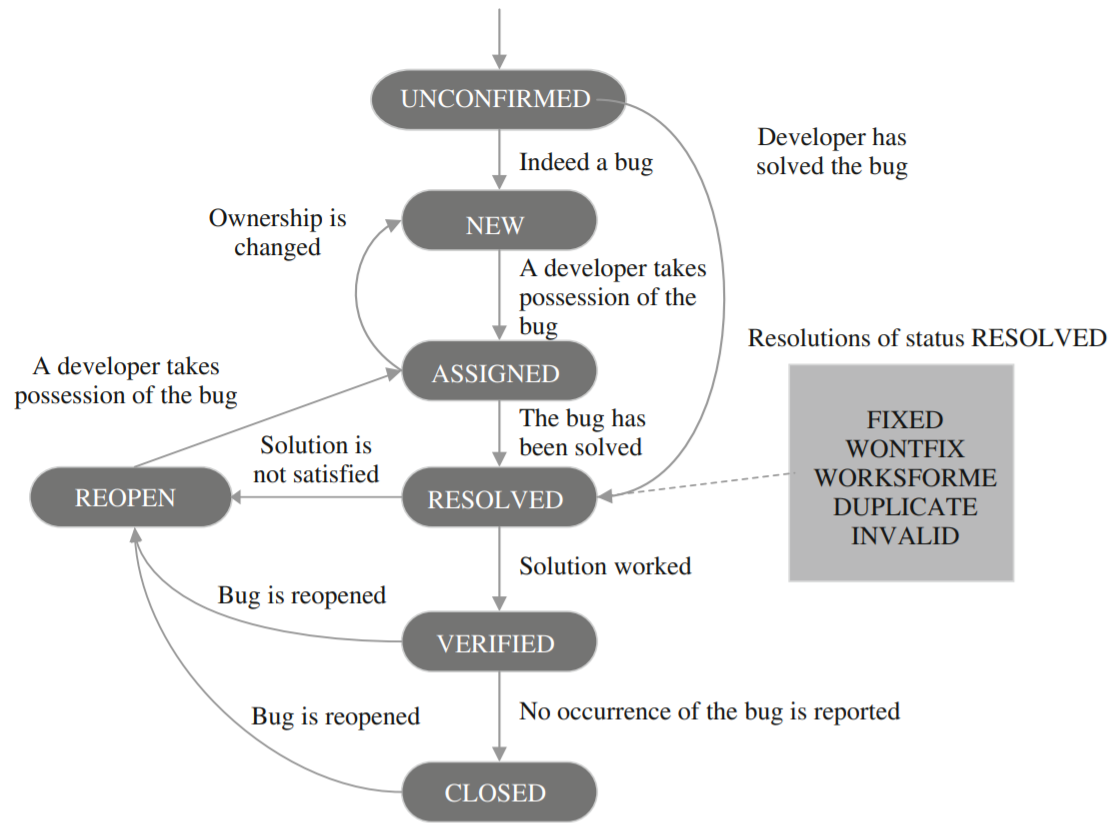}
		\caption{The life-cycle of a bug report in Bugzilla~\citep{Zhang2015}}
		\label{fig:bug-life-cycle}
	\end{figure}

	A bug report may visit different stages in its life-cycle. 
	Figure~\ref{fig:bug-life-cycle} shows the life-cycle of bug reports in Bugzilla-based projects. 
	When a bug is introduced to a bug tracking system, it is UNCONFIRMED until a developer verifies its validity. 
	As soon as it is verified, the status changes to NEW, and after the triaging phase, it is ASSIGNED to a proper developer by a triager, or a developer claims its possession. 
	Then, the developer starts fixing the bug, and after it is RESOLVED, they determine its resolution. 
	After setting the solution and ensuring the proposed solution is suitable, the bug will be VERIFIED and then CLOSED unless the fix is not satisfactory or the same bug happens in future releases. 
	In that case, it will be REOPENed. 
	
	A reported bug possesses different attributes, some of which are filled at reporting date while others may be determined later in the fixing phase.
	Therefore, some features, e.g., bug dependency, severity, and priority, may change during the bug life-cycle, whereas others, e.g., reporter's name, title, and description, are constant. 
	Most bug triage studies disregard these changes and only highlighted static attributes -- i.e., title and description. 
	As a bug evolves, other bugs may block or be blocked by the introduced bug.
	Accordingly, bug prioritization and triage process should be adjusted to incorporate newly found dependencies. 
	In a typical bug triage task, bugs may be deemed to be independent; however, when a triager considers inherent bug dependencies, they cannot solve blocked bugs unless the blocking bug is fixed. 
	Importantly, it is not sensible to spend time fixing bugs that are blocked by others.
	Hence, the bug triage process does not take place in a vacuum. 
	
	The bug triage can be applied simultaneously with bug prioritization to give a more comprehensive picture of the process. 
	Accordingly, we may postpone triaging bugs that are infeasible to be fixed at this time step. 
	In this study, we add a new dimension to the bug triage process by considering the bug dependencies.

	\section{Methods}\label{sec:research-methodology}
	In this work, we consider three open software systems, namely, {\scshape{EclipseJDT}}, {\scshape{Mozilla Firefox}}, and {\scshape{LibreOffice}}. 
	We examine bug evolution in software repositories and its effect on triaging process. 
	To this end, we extract reported bugs' information from their Issue Tracking System, covering a decade from January 2010 to December 2019. 
	In our triaging process, we consider both textual information of the reported bugs and their dependencies. 
	We construct a bug dependency graph (BDG), where the reported bugs are its nodes, and the blocking information, i.e., ``depends on'' and ``blocks`` forms its arcs. 
	A BDG is a directed acyclic graph (DAG) in which bugs cannot block other bugs while they are simultaneously blocked by the same bugs in a loop. 
	Bug dependencies are usually identified during the fixing process; hence, we assume the BDG is constantly evolving by the changes in its nodes -- i.e., fixing a bug or introducing a new bug -- and its arcs -- i.e., finding a new dependency or removing an existing dependency.
	
	We propose Dependency Aware Bug Triaging that assigns the bug to the most appropriate developer while considering the workload of available developers and the dependency between the bugs. 
	In particular, we explore the following research questions.
	\begin{enumerate}[start=0,label={(\bfseries RQ\arabic*):}]
		\item \textbf{How can DABT prevent task concentration on developers?}
		\item \textbf{How can DABT reduce the fixing time and the overdue bugs at the same time?}
		\item \textbf{How can DABT help developers to postpone the blocked bugs?}
	\end{enumerate}
	
	\subsection{Dependency-aware bug triaging}
	As we discussed in Section~\ref{sec:existing-methods}, the textual information and the fixing time of the bugs are of importance in bug triaging automation. 
	Studies focusing on CosTriage methods claim that merely looking for the textual-information may increase the accuracy; however, it leads to longer fixing times. 
	Therefore, they define a control parameter, $\alpha$, to make a trade-off between accuracy and the fixing cost. 
	However, both CBR and CosTriage approaches ignore the impact of the imbalanced distribution of the fixing tasks among developers. 
	The release-aware bug triaging method enriches the triaging process with the predefined constraint on developers' schedules to minimize overdue bugs. 
	On the other hand, none of the previous methods consider the importance of the BDG while assigning a bug to a developer. 
	If a bug that is not fixable due to its dependency is triaged, we might waste the valuable time of the developers. 
	
	Our proposed algorithm, DABT, incorporates the main ideas from the previously proposed approaches and enhances those by explicitly modeling the bug dependency. Its assumptions can be summarized as follows.
	
	\begin{itemize}\setItemSep{0.5em}
		\item We have a fixed number of developers, $d \in \{1, 2, \dots, D\}$ working on resolving bugs by taking into account the BDG. 
		We obtain the list of developers using predefined conditions on \textit{active} developers. Specifically, we take $D$ as 28, 86, and 124 for {\scshape{EclipseJDT}}, {\scshape{LibreOffice}} and {\scshape{Mozilla}}, respectively. 
		
		\item In this study, we do not consider the severity and priority of the bugs since previous works showed that they are unreliable and subjective~\cite{jahanshahi2020, akbarinasaji2018partially}. 
		
		\item Bug $i$ has fixing time of $c_i^d$ if it is solved by developer $d$, as demonstrated in Figure~\ref{fig:BDG}. Also, $c_i$ indicates the set of cost values of bug $i$ for all active developers.
		For our analysis to be consistent and comparable with those of previous studies~\cite{kashiwa2020, park2011costriage}, we estimate the fixing time in the same manner. 
		We apply LDA topic modeling, use Arun's method to obtain the optimal number of categories, find the average fixing time of each developer given the category, and finally, estimate the missing values using a collaborative filtering recommender.
		\begin{figure}[!htb]
			\centering
			\begin{tikzpicture}[b/.style={circle,draw,execute at begin node={$b_{#1}$},
				alias=b-#1,label={[rectangle,draw=none,overlay,alias=l-#1]right:{$[s_{#1}^d,c_{#1}^d]$}}}]
			\node[matrix of nodes,column sep=1em,row sep=2em]{
				& & |[b=1]|& & |[b=2]| & & &|[b=7]|\\
				&  |[b=3]|& & |[b=4]| &  & &|[b=8]| &\\
				|[b=5]|& & |[b=6]| &  & & &|[b=9]| &\\
			};
			\path[-stealth] foreach \X/\Y in {1/3,3/5,3/6,1/4,2/4,8/9} {(b-\X) edge (b-\Y)};
			\path (l-7.east); 
			\end{tikzpicture}
			\caption{A typical BDG}
			\label{fig:BDG}
		\end{figure}
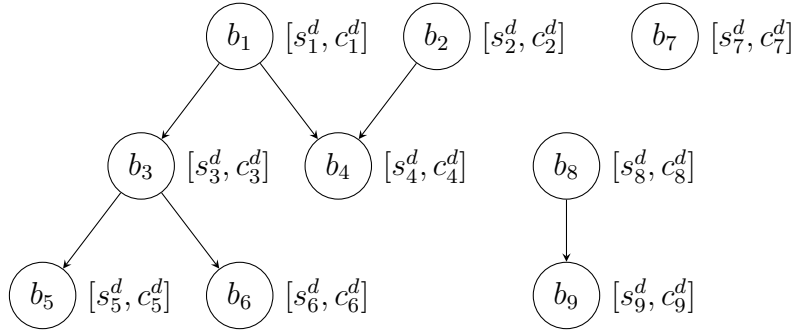
		
		Figure~\ref{fig:BDG} also demonstrates the dependencies among the bugs. 
		For instance, $b_1$ is the blocking bug for $b_3$ and $b_4$. We cannot fix those two bugs unless their parent nodes (i.e., blocking bugs) are resolved. 
		Solo bugs are very common in such graphs (e.g., see $b_7$), which neither block nor are blocked by other bugs. 
		When the number of arcs (i.e., dependencies) in the BDG increases, the impact of considering bug dependency in the bug triage process increases as well. 
		
		\item Bug $i$ has a \textit{suitability} $s_i^d$ when assigned to developer $d$. Also, $s_i$ denotes the set of suitability values of bug $i$ for all developers.
		The notion of suitability implies that, in the triage process, we should assign the bugs to the most compatible developer. 
		To calculate the suitability, we train a model on textual information (i.e., title and description) obtained from the history of resolved bugs.
		A TF-IDF converts the merged textual data to numeric values and makes those ready for the classifier. 
		We adopt an SVM classifier that incorporates the output of TF-IDF as the independent features and the developers' names as the class labels. 
		We fit the model at the end of the training phase. 
		Then, it can predict the suitability of the developers given the textual information of a new bug. 
		We adopt SVM implementation of scikit-learn in Python with a linear kernel and the regularization parameter, $C$, set to 1000. 
		These settings are compatible with previous works. 
		We use default values for other parameters. 
		Figure~\ref{fig:suitability} shows the process of $s_i^d$ estimation.
		\vspace{-0.2cm}
		\begin{figure}[!ht]
			\vspace{-0.1cm}
			\centerline{\includegraphics[width=\linewidth]{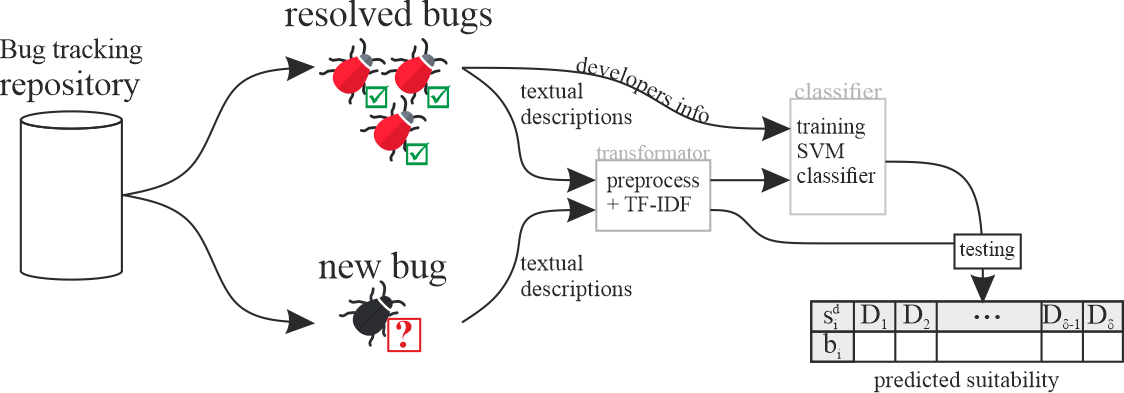}}
			\caption{Suitability estimation~\cite{kashiwa2020}}
			\label{fig:suitability}
		\end{figure}
		
		\item Similar to \citet{kashiwa2020}, we solve the IP model or other baseline methods at the end of each testing day. 
		Accordingly, developers are assigned bugs once a day.
		
		\item A single bug cannot be assigned to more than one developer at the same time.
		\item Bug dependency is updated at the moment of finding a dependency or removing one. Therefore, its updates are instant and separate from the model and are done by the simulator. 
	\end{itemize}
	
	We next provide our IP model that determines the assignment of the bugs to the developers. Let $x_i^d$ be a binary variable that takes value 1 if bug $i$ is solved by developer $d$, and 0 otherwise. In addition, let $T_t^d$ denote the time limit of developer $d$ at time $t$. Specifically, at time step $t$, developers' time limit is updated according to their schedule and previously assigned bugs. We define an identical upper limit $\mathcal{L}$ for all developers on their $T_t^d$ values. This upper limit is equivalent to the maximum planning horizon for a project and can change according to the project size. We set $\mathcal{L}$ to the third quartile value of the bug fixing times according to the previous study~\cite{kashiwa2020}. The value of $\mathcal{L}$ for  {\scshape{EclipseJDT}}, {\scshape{LibreOffice}} and {\scshape{Mozilla}} is 26, 10.5 and 18 days, respectively. Also, we use $P(i)$ to denote the list of parents of bug $i$. The IP model for bug assignment can then be formulated as follows.
	\begin{maxi!}|l|
		{}{\sum_{d} \sum_i \Big(\big(\alpha \times \frac{s_i^d}{\max(s_i)}\big) + \big( (1-\alpha) \times \frac{\nicefrac{1}{c_i^d}}{\nicefrac{1}{\min(c_i)}} \big) \Big) x_i^d  \label{eq:IP-objective}}{}{}
		\addConstraint{x_i^d}{\leq x_p^d}{\quad \forall d, \ \  \forall i \neq \text{root}, p \in P(i) \label{eq:IP-1}}
		\addConstraint{\sum_i c_i^d x_i^d}{\leq T_t^d }{\quad\forall d }\label{eq:IP-2}
		\addConstraint{\sum_{d} x_i^d}{\leq 1}{\quad\forall i \label{eq:IP-3}}
		\addConstraint{x_i^d}{\in \{0,1\}}{\quad\forall i \label{eq:IP-4}}
	\end{maxi!}
	
	The objective function of the model, \eqref{eq:IP-objective}, maximizes the suitability of the solved bugs while reducing the cost (i.e., solving time). The trade-off between suitability and estimated fixing time is determined by $\alpha$. Different from \citet{kashiwa2020}'s model, we incorporate the fixing time in the objective function since ignoring the solving time might cause over-specialization. It also helps to reduce bug fixing time. Constraints~\eqref{eq:IP-1} ensure that the precedence constraints are imposed, that is, blocking bugs $x_p^d$ are solved before the blocked bug $x_i^d$. Constraints~\eqref{eq:IP-2} enforce total time limit requirements. Lastly, constraints~\eqref{eq:IP-3} guarantee that a single bug cannot be assigned to more than one developer simultaneously.
	
	The proposed solution approach incorporates all the previously suggested criteria while enhancing those by accounting for bug dependency considerations. 
	In addition, it uses IP formulation for bug triaging that had been only considered by~\citet{kashiwa2020}. 
	Note that their formulations were different than ours in that they disregarded the effect of the bug fixing time in their objective function and did not include the dependency constraint.
	
	\subsection{Bug triaging mechanism}
	We design a mechanism to recreate all the past events.
	Unlike the previous studies that worked on the static datasets (typically stored in CSV or JSON files), we reconstruct the history of events and apply each algorithm in real-time. 
	The online exploration enables us to examine the outcome of bug assignments given other elements in the system.
	Therefore, we build a pipeline of the past events and replace the assignment task with our proposed algorithm. 
	The flow and time of the bug reports, bug reopening, and bug dependency discovery remain the same. 
	We adopt the suggested pipeline by~\cite{jahanshahi2020} to recreate past events.
	
	We first extract bug fixing information from Bugzilla. 
	Then, we revisit past events in the system during the training time and keep track of the information related to active developers and feasible bugs. 
	When the training phase is completed, SVM and LDA learn the suitability and cost list accordingly. 
	Afterwards, we continue recording the incoming bugs and their possible dependencies during the testing phase. 
	Once in a day, we query the currently feasible bugs' information in the system and apply our model to those. 
	The outcome of the model is a list of developers and assigned bugs. 
	A few bugs may remain unassigned based on the constraint we have. 
	We call back those bugs in the upcoming days until they meet the criteria. The steps of the full procedure is provided below.
	
	\begin{enumerate}[start=0,label={(\bfseries Step \arabic*):}, leftmargin = 4em]
		\item \textbf{Setting hyper-parameters}: We use the third quartile as the value for upper limit $\mathcal{L}$. All $T_t^d$ are initialized by $\mathcal{L}$ and cannot exceed this limit during the process. Furthermore, SVM and LDA hyperparameters are defined based on the previous studies. 
		
		\item \textbf{Constructing SVM and LDA}: To estimate the suitability $s_i^d$ and the fixing time $c_i^d$ for developer $d$ and bug $i$, we train SVM and LDA at the end of the training phase. 
		
		\item \textbf{Predicting suitability and costs}: At the end of each day, we predict the suitability and cost of each bug for all feasible, unassigned bugs. 
		
		\item \textbf{Applying multiple knapsack problem}: By solving the IP model, we determine the bugs and their associated developers. 
		Based on the constraints, the model postpones certain bugs while prioritizing others.
		
		\item \textbf{Updating $T_t^d$}: At the end of each day, we increment $T_t^d$ by 1 for developer $d$ while ensuring it does not exceed $\mathcal{L}$. 
		At the same time, we decrease $T_t^d$ of developer $d$ by sum of their estimated $c_i^d$ values for all assigned bugs.
		
		\item \textbf{Continuing to the next day (to step 2)}: Once the assignment at day $t$ finishes, we move to the next day and repeat the process from step 2. 
		These steps will continue until the end of the testing phase. 
	\end{enumerate}
	
	We assume that our method and baselines are implemented once a day. 
	However, the process is generalizable, and the granularity of the updating time can be modified as needed.
	
	\subsection{Dataset}
	We consider three large open-source software (OSS) projects in our analysis. 
	They are well-established projects with a sufficient number of bug reports. 
	These projects were also adopted in the previous studies~\citep{Bhattacharya2010, lee2017applying, mani2019deeptriage, kashiwa2020}, making them suitable for comparison. 
	We collect the raw data from the repository using the Bugzilla REST API~\footnote{\hyperlink{https://wiki.mozilla.org/Bugzilla:REST_API}{https://wiki.mozilla.org/Bugzilla:REST\_API}}. 
	It includes general attributes of the bugs and their metadata change history. 
	Table~\ref{tab:dataset} shows the information of the extracted datasets. 
	We use bug reports for the past two years as the testing set and the older ones as the training set. 
	
	\begin{table*}[!ht]
		\caption{The information of the extracted datasets. The training phase starts from Jan. 1st, 2010 to Dec. 31st, 2017, whereas the testing phase covers Jan. 1st, 2018 to Dec. 31st, 2020.\label{tab:dataset}}%
		\resizebox{\linewidth}{!}{
			\begin{tabular}{lrrrrrrrrr}
				\toprule
				& \multicolumn{3}{c}{{\scshape{\textbf{EclipseJDT}}}} & \multicolumn{3}{c}{{\scshape{\textbf{LibreOffice}}}} & \multicolumn{3}{c}{{\scshape{\textbf{Mozilla}}}} \\ 
				& \textbf{Training} & \textbf{Testing} & \textbf{Total} & \textbf{Training} & \textbf{Testing} & \textbf{Total} & \textbf{Training} & \textbf{Testing} & \textbf{Total} \\
				\midrule
				\textbf{Total bugs reported} & 12,598 & 3,518 & 16,116 & 55,554 & 14,582 & 70,136 & 90,178 & 22,353 & 112,531 \\
				\textbf{Total bug dependencies found} & 2,169 & 970 & 3,139 & 28,472 & 21,894 & 50,366 & 71,549 & 19,223 & 90,772 \\
				\textbf{Total relevant changes in the bugs' history} & 55,109 & 15,505 & 70,614 & 267,310 & 94,682 & 361,992 & 410,010 & 114,778 & 524,788 \\
				\textbf{Mean and Median fixing time (days)} & (39.1, 3) & (15.7, 1) & (33.4, 2) & (35.5, 2) & (9.4, 2.) & (29.3, 2) & (31.3, 5) & (12.6, 4) & (26.0, 5) \\
				\textbf{Minimum, Maximum fixing time (days)} & (1, 1,753) & (1, 423) & (1, 1,753) & (1, 1,509) & (1, 428) & (1, 1,509) & (1, 2,610) & (1, 550) & (1, 2,610) \\ 
				\midrule
				\textbf{After cleaning} &  &  &  &  &  &  &  &  &  \\
				\qquad \textbf{1. Bugs with resolved status} & 11,146 & 2,562 & 13,708 & 46,890 & 11,106 & 57,996 & 79,281 & 18,742 & 98,023 \\
				\qquad \textbf{2. Bugs assigned to active developers} & 5,212 & 1,483 & 6,695 & 6,569 & 1,890 & 8,459 & 15,322 & 6,403 & 21,725 \\
				\qquad \textbf{3. Bugs with known assignment date} & 4,193 & 1,348 & 5,541 & 5,655 & 1,754 & 7,409 & 10,004 & 3,960 & 13,964 \\
				\qquad \textbf{4. Bugs with acceptable fixing time} & 3,598 & 1,251 & 4,849 & 4,708 & 1,570 & 6,278 & 8,651 & 3,704 & 12,355 \\
				\bottomrule
			\end{tabular}
		}
	\end{table*}
	
	Similar to previous studies~\citep{kashiwa2020,park2011costriage}, we consider only the bugs that meet four criteria.
	\begin{itemize}\setItemSep{0.5em}
		\item Not all bugs have fixing time information available. 
		For instance, some bugs are still open, or in some cases, the exact fixing date is not available in history. 
		Hence, we only consider the FIXED or CLOSED bugs with sufficient information.
		
		\item The assignment date is not available for some bugs, or the assignment time is after their resolution. 
		We eliminate them as invalid bugs.
		
		\item Some bugs took years to get solved. 
		We remove outlier bugs whose fixing time is greater than the threshold of Q3 + (1.5 $\times$ IQR), where Q3 is the third quartile of the fixing time, and IQR is the interquartile range, i.e., 
		\begin{equation*}
			\text{IQR} = (\text{the third quartile}) - (\text{the first quartile}).
		\end{equation*}
		
		The maximum acceptable fixing times that we found for {\scshape{EclipseJDT}}, {\scshape{LibreOffice}} and {\scshape{Mozilla}} are 63.5, 24.8 and 42 days, respectively.
		
		\item Similar to previous works, we consider only \textit{active} developers. 
		We exclude inactive developers as we do not have sufficient information on them. 
		Considering IQR as a measure of central distribution, we define \textit{active} developers as the ones whose bug fix number is higher than the IQR of bug fix numbers of each developer. 
	\end{itemize}
	
	Table~\ref{tab:dataset} shows the effect of applying each filtering step on the number of feasible bugs in the system.
	We preprocess the textual information through lemmatization, stop words, numbers and punctuation removal, and lengthy word elimination (i.e., longer than 20 characters). 
	Finally, we merge titles and descriptions of the bugs and tokenize them.

	\section{Evaluation}~\label{sec:results}
	We compare the proposed method, DABT, with four other alternatives, i.e., the actual bug assignment, CBR using SVM, CosTriage, and RABT. We define different metrics for the comparison as shown in Table~\ref{tab:comp_algorithms}. As the table indicates, RABT and DABT have few unassigned bugs at the end of the testing phase due to the constraints they impose. This number is negligible compared to the total bugs that are addressed. We implemented all the methods in Python and used Gurobi 9.0 to solve the IP models.
	
	\begin{table*}[!ht]
		\caption{Comparison of different algorithms\label{tab:comp_algorithms}}%
		\resizebox{\linewidth}{!}{
			\begin{tabular}{cl>{\columncolor[HTML]{EFEFEF}}r rrrr} 
				\toprule
				& \textbf{Metrics} & \textbf{Actual} & \textbf{CBR} & \textbf{CosTriage} & \textbf{RABT} & \textbf{DABT} \\
				\midrule
				\multirow{11}{*}{\STAB{\rotatebox[origin=c]{90}{\scshape{\textbf{EclipseJDT}}}}} 
				& \textbf{\# of assigned bugs} & 1,250 & 1,250 & 1,250 & 1,244 & 1,238 \\
				& \textbf{\# of un-assigned bugs} & 0 & 0 & 0 & 6 & 12 \\
				& \textbf{\# of assigned developers} & 15 & 19 & 19 & 28 & 27 \\
				& \textbf{Task distribution among developers $(\mu, \sigma)$} & (83.4, 93.7) & (65.8, 112.0) & (65.8, 108.5) & (\textbf{44.4}, \textbf{39.0}) & (45.9, 52.1) \\
				& \textbf{Mean Fixing days per bug} & 6.0 & 7.9 & 7.5 & 7.0 & \textbf{4.4} \\
				& \textbf{Percentage of overdue bugs} & 66.0 & 82.2 & 79.6 & 17.6 & \textbf{11.9} \\
				& \textbf{Percentage of un-fixed bugs} & 66.0 & 82.2 & 79.6 & 18.1 & \textbf{12.9} \\
				& \textbf{Accuracy of assignments} & 97.7 & \textbf{95.5} & 94.0 & 83.4 & 73.2 \\
				& \textbf{Infeasible assignment w.r.t. bug dependency} & 5.4 & 6.0 & 5.8 & 4.7 & \textbf{0.0} \\
				& \textbf{Mean Depth of the BDG} & 0.05 & 0.05 & 0.05 & 0.04 & 0.04 \\
				& \textbf{Mean Degree of the BDG} & 0.05 & 0.05 & 0.05 & 0.04 & 0.04 \\
				\midrule
				\multirow{11}{*}{\STAB{\rotatebox[origin=c]{90}{\scshape{\textbf{LibreOffice}}}}} 
				& \textbf{\# of assigned bugs} & 1,569 & 1,569 & 1,569 & 1,569 & 15,68 \\
				& \textbf{\# of un-assigned bugs} & 0 & 0 & 0 & 0 & 1 \\
				& \textbf{\# of assigned developers} & 57 & 22 & 21 & 86 & 49 \\
				& \textbf{Task distribution among developers $(\mu, \sigma)$} & (27.5, 68.9) & (71.3, 224.5) & (74.7, 253.2) & (\textbf{18.2}, \textbf{59.3}) & (32.0, 78.7) \\
				& \textbf{Mean Fixing days per bug} & 3.3 & 2.1 & 1.8 & 2.6 & \textbf{1.6} \\
				& \textbf{Percentage of overdue bugs} & 35.9 & 77.1 & 80.8 & 15.6 & \textbf{13.2} \\
				& \textbf{Percentage of un-fixed bugs} & 35.9 & 77.1 & 80.8 & 15.6 & \textbf{13.3} \\
				& \textbf{Accuracy of assignments} & 91.7 & 99.1 & \textbf{99.3} & 94.4 & 92.6 \\
				& \textbf{Infeasible assignment w.r.t. bug dependency} & 0.1 & 0.1 & 0.1 & 0.1 & \textbf{0.0} \\
				& \textbf{Mean Depth of the BDG} & 1.41 & 1.41 & 1.41 & 1.41 & 1.41 \\
				& \textbf{Mean Degree of the BDG} & 0.84 & 0.84 & 0.84 & 0.84 & 0.84 \\
				\midrule
				\multirow{11}{*}{\STAB{\rotatebox[origin=c]{90}{\scshape{\textbf{Mozilla}}}}} 
				& \textbf{\# of assigned bugs} & 3,704 & 3,704 & 3,704 & 3,704 & 3,687 \\
				& \textbf{\# of un-assigned bugs} & 0 & 0 & 0 & 0 & 16 \\
				& \textbf{\# of assigned developers} & 137 & 74 & 85 & 124 & 114 \\
				& \textbf{Task distribution among developers $(\mu, \sigma)$} & (27.0, 49.5) & (50.1, 204.0) & (43.6, 187.0) & (\textbf{29.9}, \textbf{39.5}) & (32.3, 59.3) \\
				& \textbf{Mean Fixing days per bug} & 7.0 & 7.2 & 6.6 & 7.0 & \textbf{3.3} \\
				& \textbf{Percentage of overdue bugs} & 69.8 & 80.1 & 77.6 & 25.0 & \textbf{12.7} \\
				& \textbf{Percentage of un-fixed bugs} & 69.8 & 80 & 77.6 & 25.0 & \textbf{13.2} \\
				& \textbf{Accuracy of assignments} & 72.7 & \textbf{60.2} & 59.0 & 50.9 & 33.2 \\
				& \textbf{Infeasible assignment w.r.t. bug dependency} & 9.4 & 9.0 & 8.8 & 6.8 & \textbf{0.0} \\
				& \textbf{Mean Depth of the BDG} & 0.42 & 0.44 & 0.43 & \textbf{0.39} & \textbf{0.39} \\
				& \textbf{Mean Degree of the BDG} & 0.27 & 0.27 & 0.27 & \textbf{0.25} & \textbf{0.25} \\
				\bottomrule
			\end{tabular}
		}
	\end{table*}
	
\begin{figure}[!ht]
		\centering
		\begin{subfigure}[b]{\linewidth}
			\centering
			\includegraphics[width=0.6\textwidth]{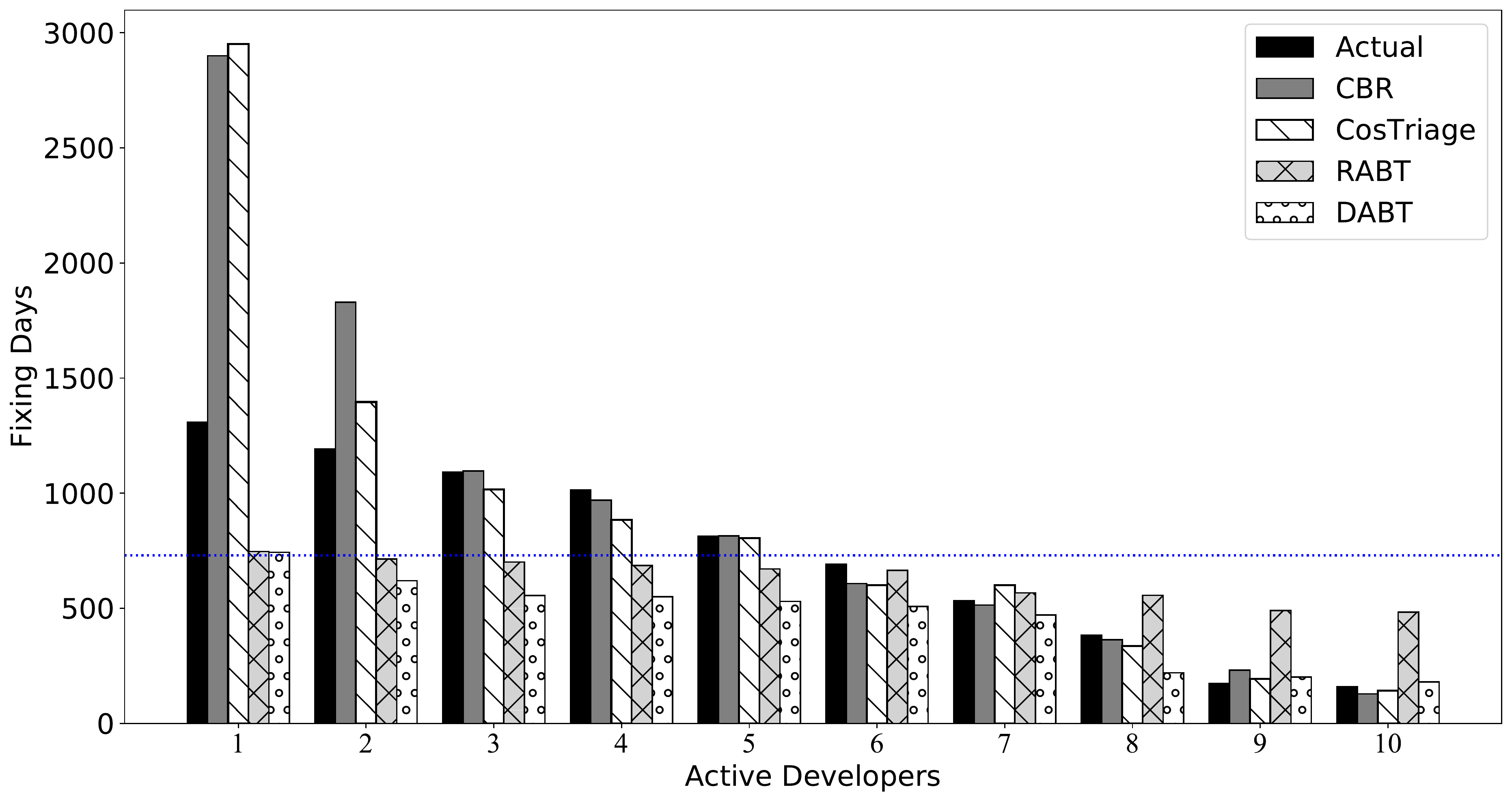}
			\caption{{\scshape{EclipseJDT}}}
			\label{fig:developer_eclipse}
		\end{subfigure}\\
		\begin{subfigure}[b]{\linewidth}
			\centering
			\includegraphics[width=0.6\textwidth]{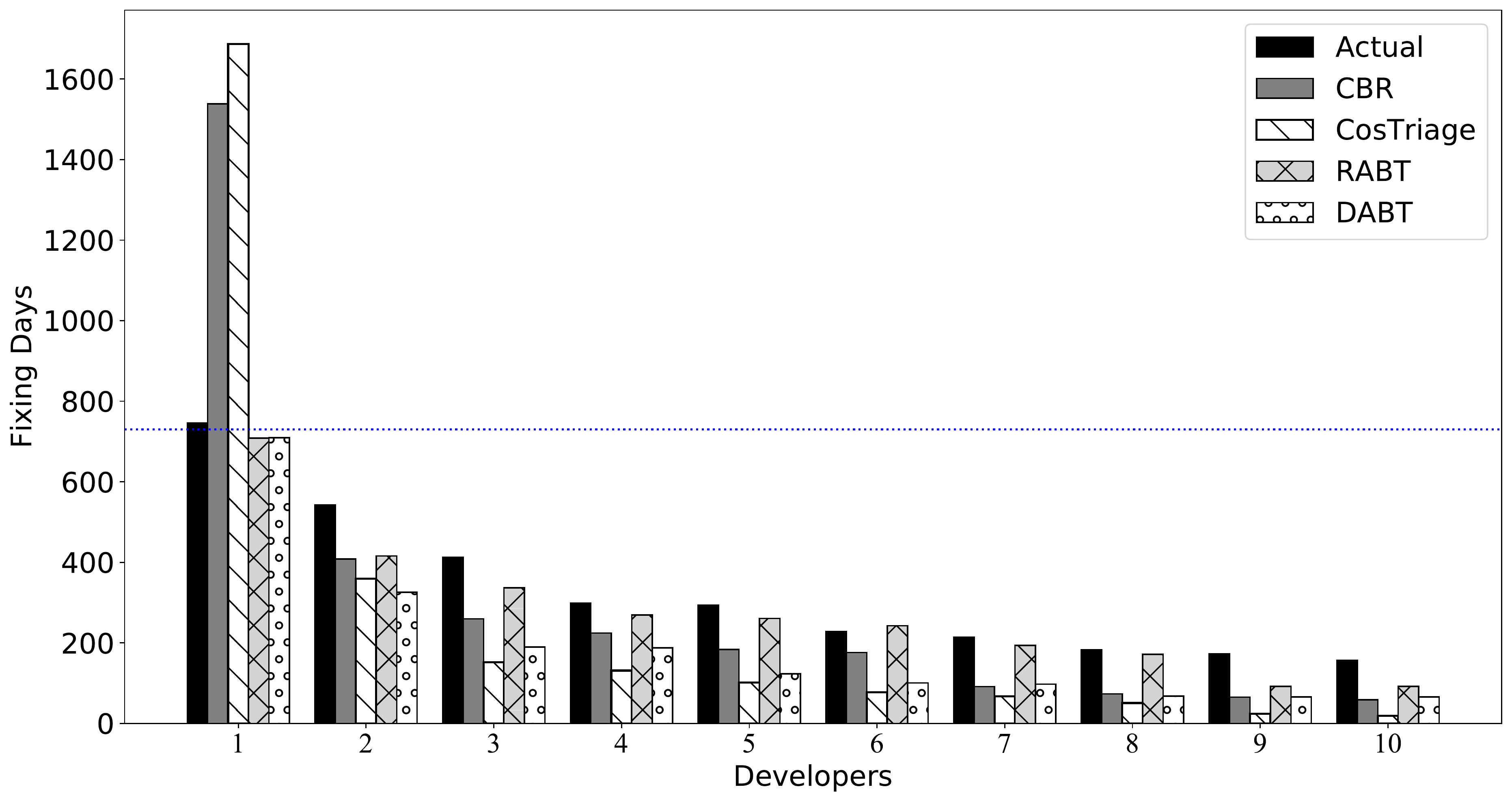}
			\caption{{\scshape{LibreOffice}}}
			\label{fig:developer_LibreOffice}
		\end{subfigure} \\
		\begin{subfigure}[b]{\linewidth}
			\centering
			\includegraphics[width=0.6\textwidth]{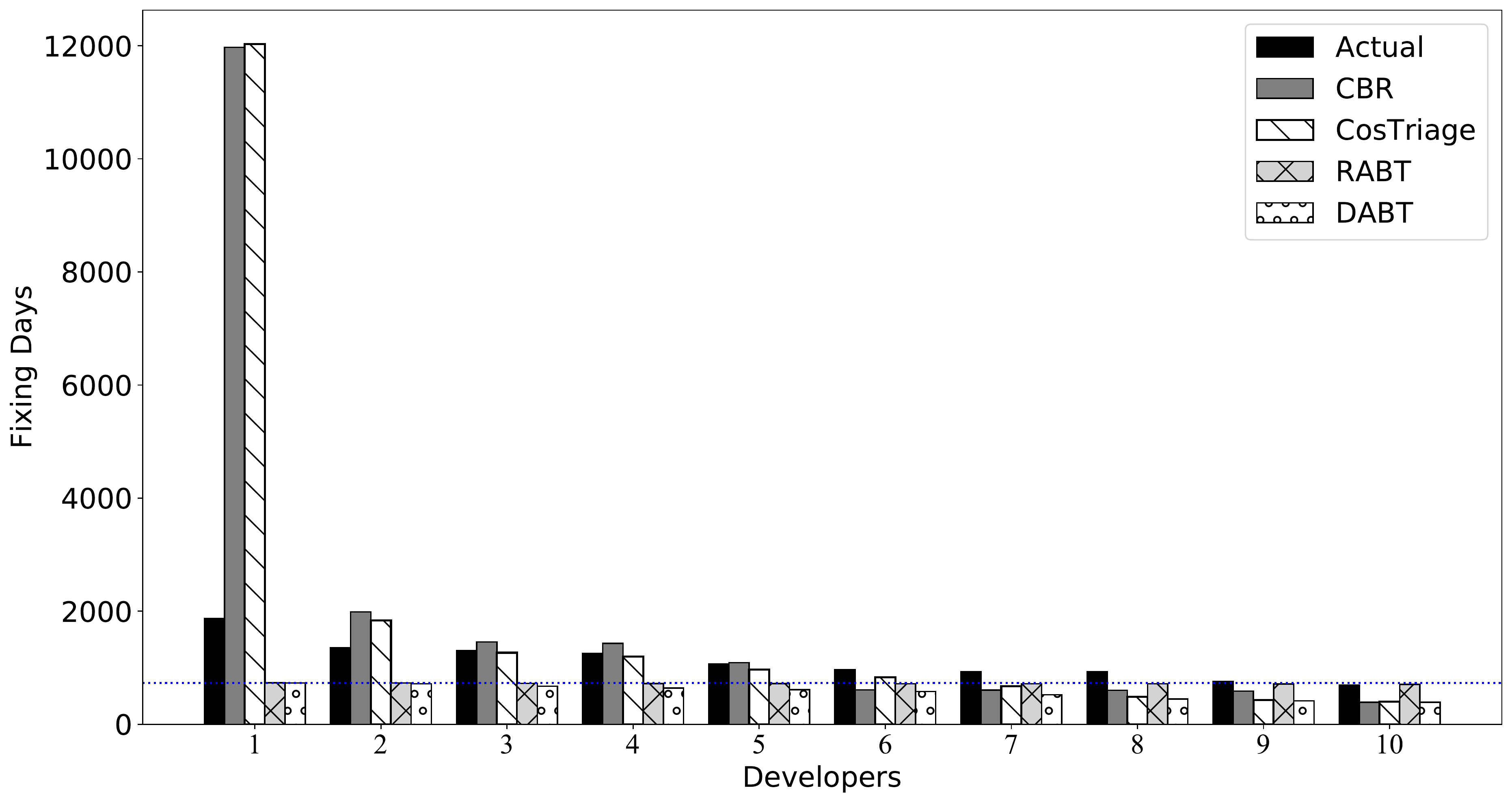}
			\caption{{\scshape{Mozilla}}}
			\label{fig:developer_Mozilla}
		\end{subfigure} 
		\caption{The number of fixing days spent by top-10 active developers during the two-year testing phase.}
		\label{fig:developer_distribution}
	\end{figure}

	One of the main differences between the algorithms is in their use of available developers. 
	Considering two bigger projects, {\scshape{LibreOffice}} and {\scshape{Mozilla}}, we observe that actual assignment, RABT, and DABT use more developers than the other two methods. 
	The similarity between the manual assignment and RABT/DABT indicates that all consider developers' schedules. 
	CBR and CosTriage do not incorporate task concentration in their formulation, and they are still prone to over-specialization. 
	We explore this fact further by reporting the average and standard deviation of the number of bugs assigned among developers. 
	Both RABT and DABT maintain the level of fair bug distribution among developers by calling more developers. 
	DABT performs slightly worse in terms of distribution compared to RABT; however, it achieves a comparable result using a smaller number of developers during the testing phase. It means that it does not call redundant developers to achieve that fair distribution. 
	We further investigate the effect of task concentration by looking for top-10 active developers of each algorithm.
	Figure~\ref{fig:developer_distribution} shows the number of fixing days for each developer. 
	As CBR and CosTriage do not consider the bug fixing loads, they even assigned bugs to their top developer up to 16 times their capacity (see Figure~\ref{fig:developer_Mozilla}). 
	Although both RABT and DABT never over-assign bugs during the two-year testing time, DABT significantly decreases the fixing time for each developer by adding fixing time minimization to its objective function. 
	Accordingly, our method both meets the schedule criterion and optimizes the fixing time. 
	
	\smallskip
	\noindent\fcolorbox{black}{white}{%
		\minipage[t]{\dimexpr1\linewidth-2\fboxsep-2\fboxrule\relax}
		\textit{\textbf{RQ1-} In general, RABT is able to alleviate the task concentration of the developers.
			Not only does it decrease the workload of expert developers, but it also reduces the total working time of developers through fixing time consideration.}
		\endminipage}
	\smallskip
	
	Table~\ref{tab:comp_algorithms} shows that the average fixing days of the bugs significantly decreases when using DABT. 
	It consistently has a lower fixing duration for different projects. 
	It justifies the importance of the tweak in the objective function of DABT that prioritizes the bugs with shorter fixing times during bug assignments. 
	Moreover, DABT has the lowest rate of overdue bugs. 
	When considering accumulated fixing time in the bug triage process, many bugs remain unresolved due to task concentration on particular developers. 
	It justifies the reason why CBR and CosTriage have a high percentage of overdue bugs and is compatible with Figure~\ref{fig:developer_distribution}. 
	By setting the hyperparameter $\mathcal{L}$, we do not allow the system to assign bugs more than the developers' capacity. 
	On the other hand, unlike RABT, we also emphasize the significance of the bug fixing time, leading to a lower rate of overdue bugs. 
	To further decrease the ratio of delayed bugs, we recommend triagers setting a dynamic value for $\mathcal{L}$, updated by the remaining time until the next release.
	
	\smallskip
	\noindent\fcolorbox{black}{white}{%
		\minipage[t]{\dimexpr1\linewidth-2\fboxsep-2\fboxrule\relax}
		\textit{\textbf{RQ2-} RABT significantly decreases the ratio of overdue bugs by considering an upper limit on the project horizon. 
			It further improves this ratio by assigning bugs with a shorter fixing time. 
			Hence, it leads to a smoother bug fixing process and addressing more bugs before the release date.}
		\endminipage}
	\smallskip
	
	Similar to previous works~\citep{kashiwa2020,park2011costriage}, we define an accurate assignment as recommending a bug to a developer who has experience in the same component. 
	Therefore, a proper bug assignment does not mean assigning to the same developer. 
	According to this definition, even the manual bug assignment case fails to achieve an accuracy of 100\% when compared to the training phase. 
	Some developers attempt to address a bug of a new component for the first time. The component might be similar but not the same as the previous ones. 
	Hence, even the actual assignment achieves an accuracy of 72.7\% for {\scshape{\textbf{Mozilla}}}. 
	It shows that expecting a high accuracy of the model may cause incorrect assignment based on the ground truth values.
	DABT has a lower accuracy for all models compared to its counterparts. 
	There is a trade-off between the accuracy and speed of the bug fixing process. 
	The parameter $\alpha$ in our model regulates the accuracy of the assignments and speed of the bug fixing. 
	Moreover, we impose an upper limit on the project time horizon that may reduce the accuracy but leads to fewer overdue bugs. 
	We further investigate the model sensitivity to its parameter to see how much of its lower accuracy arises from the regulation parameter.
	
\begin{figure*}[!ht]
		\centering
		\begin{subfigure}[b]{0.31\textwidth}
			\includegraphics[width=\textwidth]{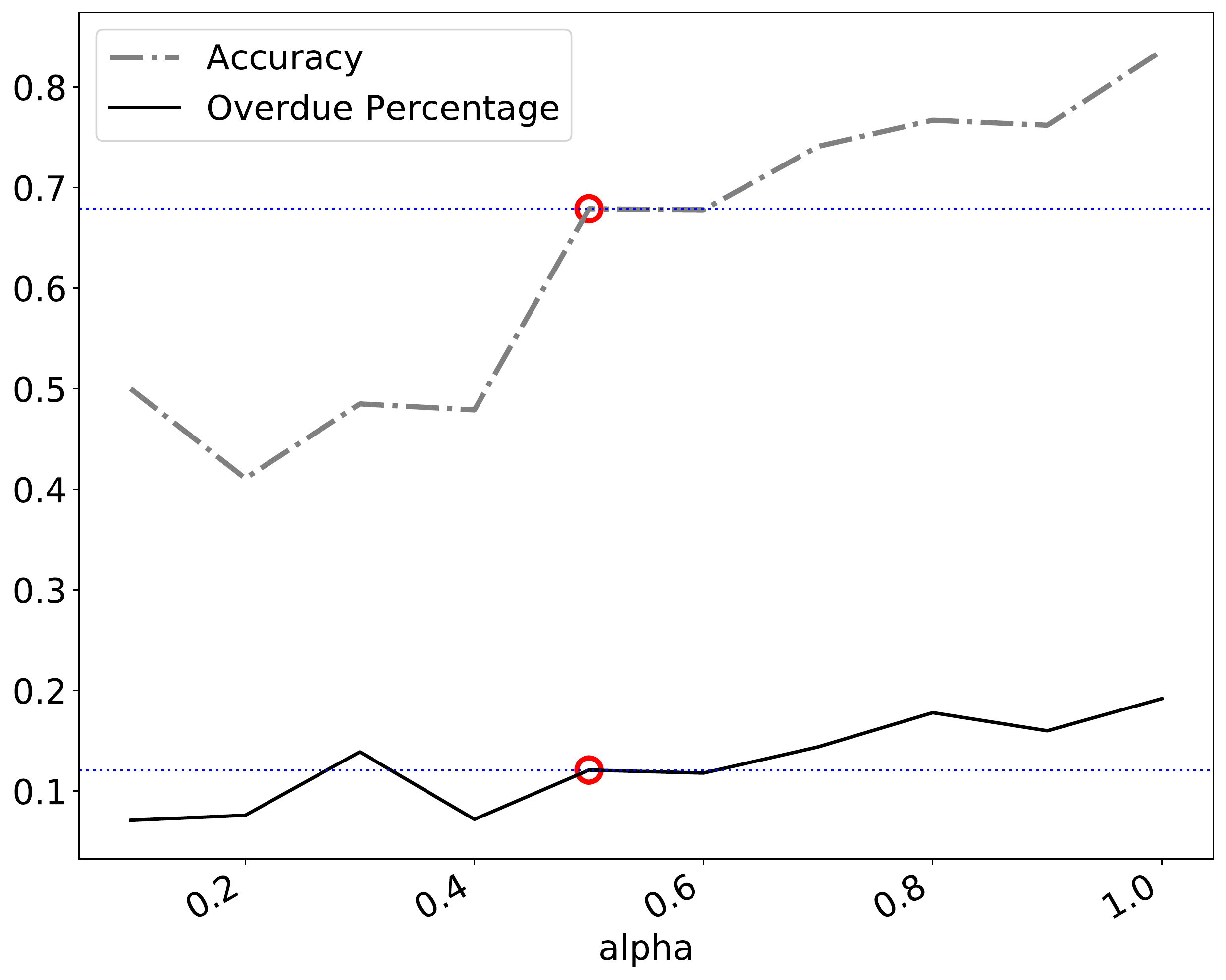}
			\caption{{\scshape{EclipseJDT}}}
			\label{fig:alpha_eclipse}
		\end{subfigure}
		\begin{subfigure}[b]{0.31\textwidth}
			\includegraphics[width=\textwidth]{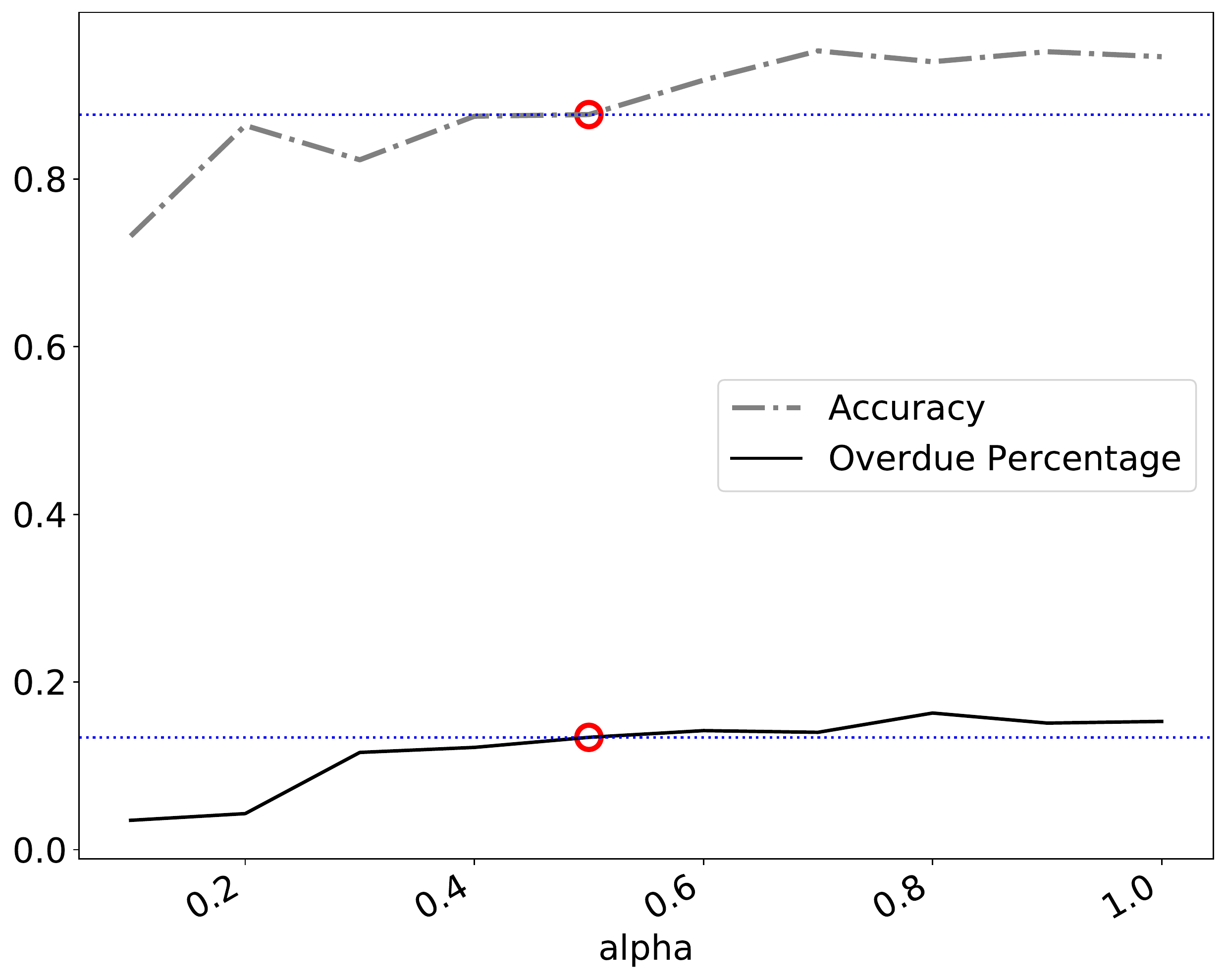}
			\caption{{\scshape{LibreOffice}}}
			\label{fig:alpha_LibreOffice}
		\end{subfigure} 
		\begin{subfigure}[b]{0.31\textwidth}
			\includegraphics[width=\textwidth]{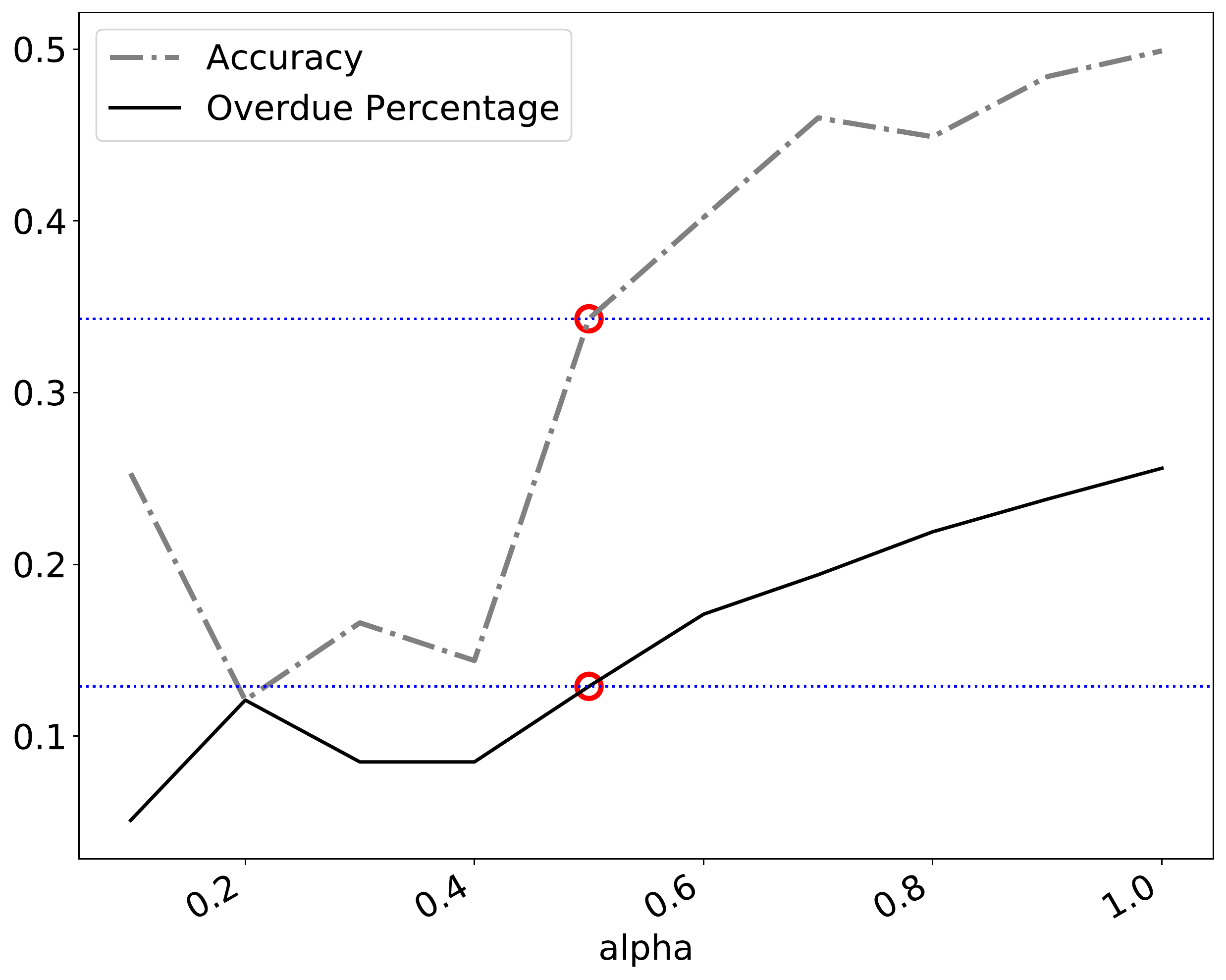}
			\caption{{\scshape{Mozilla}}}
			\label{fig:alpha_Mozilla}
		\end{subfigure} 
		\caption{Sensitivity of the accuracy and the percentage of overdue bugs of DABT to its parameter $\alpha$.}
		\label{fig:sensitivity}
	\end{figure*}

	Figure~\ref{fig:sensitivity} shows the effect of changes in $\alpha$ on the accuracy and the percentage of overdue bugs.
	As we increase $\alpha$, the model tends to give a higher weight to the developers' suitability than the bug fixing cost. 
	Therefore, DABT gets more accurate while disregarding the fixing time. 
	The developer may decide on how much accuracy they want, and accordingly, they can set a proper value for $\alpha$.
	Similar to the previous studies, we use the $\alpha = 0.5$ and give the same weight to the bug fixing time and the accuracy. 
	We believe increasing the accuracy while ignoring the associated cost may result in over-specialization and task-concentration. 
	Figure~\ref{fig:sensitivity} shows that DABT can improve its accuracy by 15\% through compromising its ability to reduce the ratio of overdue bugs. 

\begin{figure*}[!ht]
		\centering
		\begin{subfigure}[b]{0.31\textwidth}
			\includegraphics[width=\textwidth]{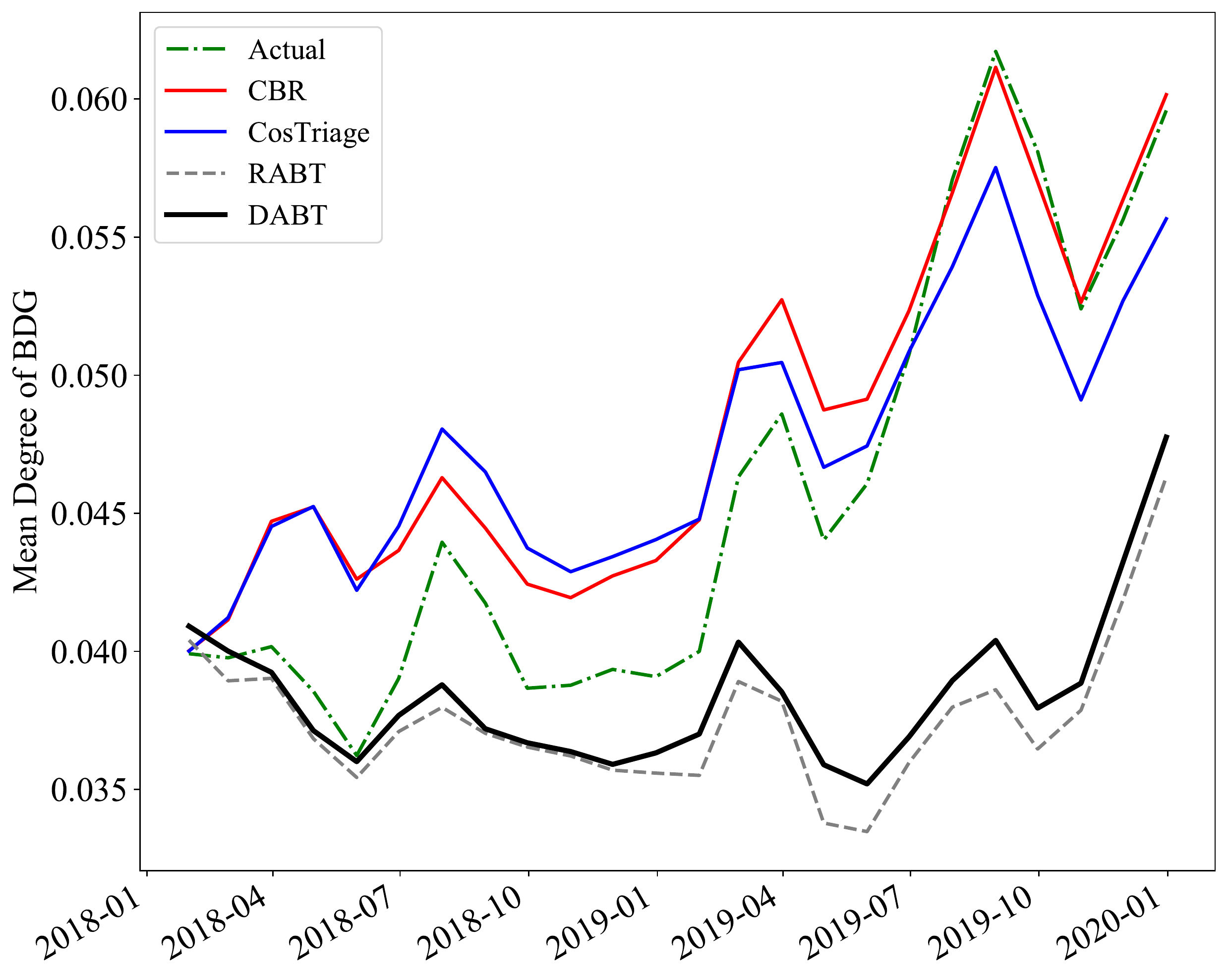}
			\caption{{\scshape{EclipseJDT}}}
			\label{fig:degree_eclipse}
		\end{subfigure}
		\begin{subfigure}[b]{0.31\textwidth}
			\includegraphics[width=\textwidth]{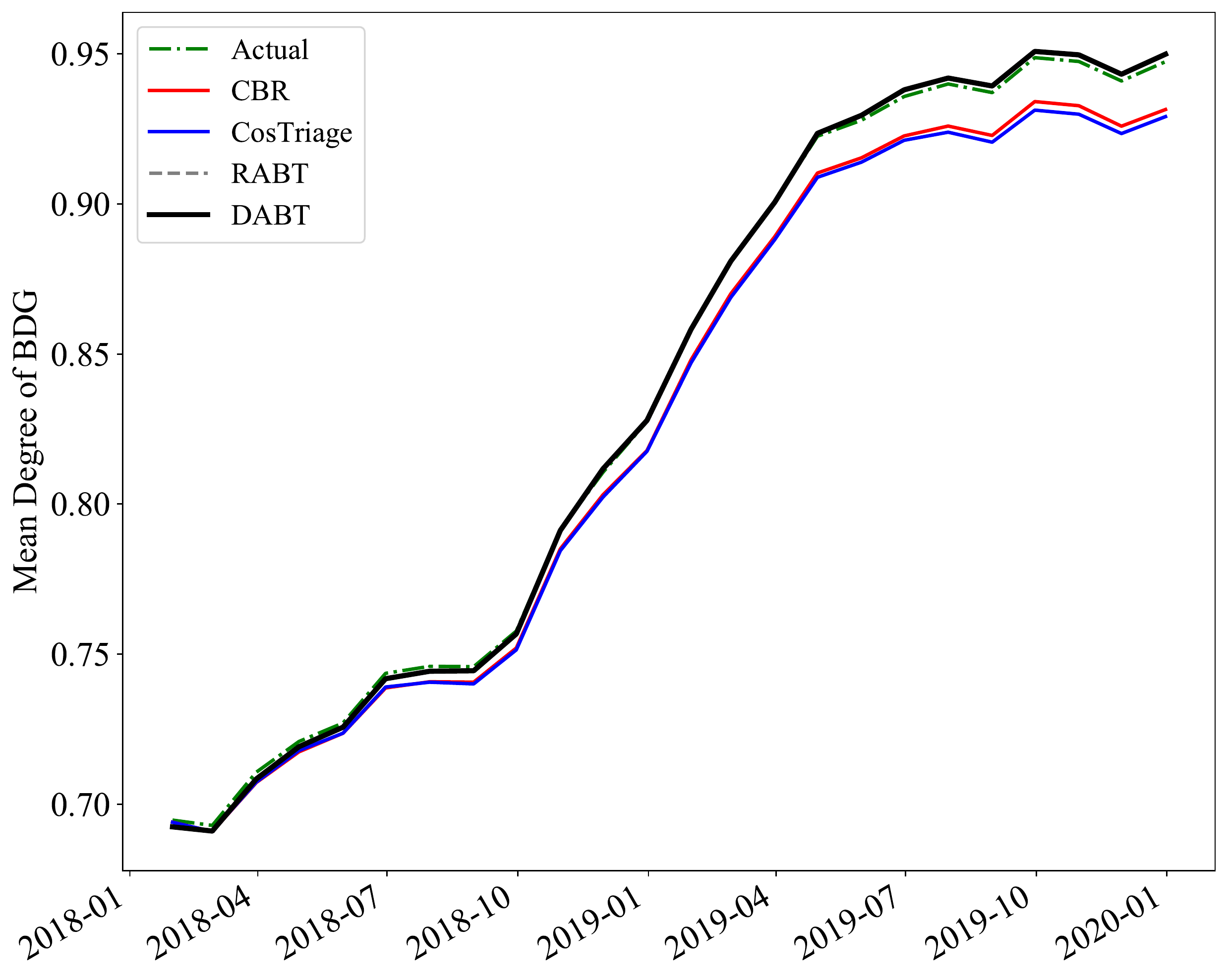}
			\caption{{\scshape{LibreOffice}}}
			\label{fig:degree_LibreOffice}
		\end{subfigure} 
		\begin{subfigure}[b]{0.31\textwidth}
			\includegraphics[width=\textwidth]{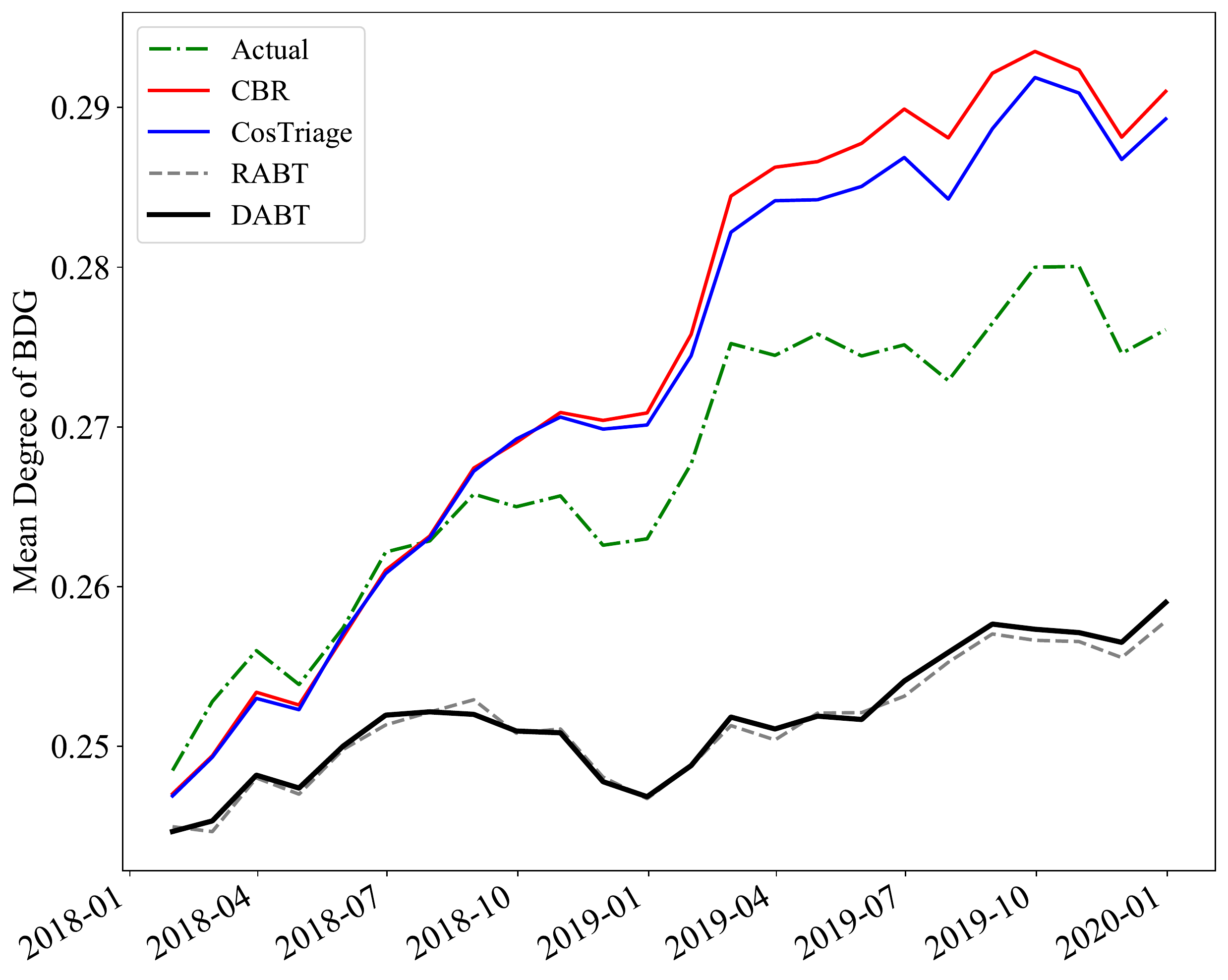}
			\caption{{\scshape{Mozilla}}}
			\label{fig:degree_Mozilla}
		\end{subfigure} 
		\caption{The effect of different strategies on the degree of the BDG during the testing phase.}
		\label{fig:degree}
	\end{figure*}

	One of DABT's key characteristics is its capability to postpone blocked bugs. 
	Whether we assign a bug or a developer wants to take possession of a bug from a lengthy list of open bugs, prioritizing the bugs that block others and increase BDG complexity is prudent. 
	Therefore, DABT includes a constraint on the infeasible bug assignment with respect to bug dependency. 
	Consequently, it postpones bugs until their blocking bugs are resolved. 
	Table~\ref{tab:comp_algorithms} shows the better performance of DABT in terms of addressing blocked bugs. 
	It also reduces the complexity of the BDG, i.e., its mean degree and depth. 
	Surprisingly, RABT also shows similar performance in terms of graph complexity. 
	Figure~\ref{fig:degree} explores the average degree of the bugs in the BDG during the two-year testing phase. We note that the mean degree of the BDG is already low since there are many solo bugs in the BDG. Nonetheless, a small reduction in these values is a significant result since it can be considered as eliminating few high-degree bottlenecks in the ITS.
	Both RABT and DABT keep the degree and the depth of the graph low, given the fact that all algorithms solve the same list of bugs. 
	For the exceptional case of {\scshape{LibreOffice}}, our finding is consistent with that of \citet{jahanshahi2020} in which they report a significant rate of found dependencies after 2017. 
	Hence, based on the other two projects, we conclude that even when addressing the same bugs, the proper timing will reduce both the complexity of the BDG and the number of overdue bugs. 
	A lower BDG complexity is beneficial in the long-run when the rate of incoming bugs is increasing while many bugs are still blocked by older ones. 
	
	\smallskip
	\noindent\fcolorbox{black}{white}{%
		\minipage[t]{\dimexpr1\linewidth-2\fboxsep-2\fboxrule\relax}
		\textit{\textbf{RQ3-} DABT is able to reduce the complexity of the bug dependency graph through the proper timing of bug assignment. It will mitigate the risk of having a high number of blocking bugs in the long-run.}
		\endminipage}
	\smallskip
	
	\section{Threats to validity}~\label{sec:threats}
	In this section, we discuss threats to the validity of our empirical study.
	
	\subsection{Construct validity}
	We estimate the performance of the models using a train-test split. 
	The first eight years are adopted as the training set and the last two years as the test. 
	However, the evolving nature of the bug repository may have an impact on the results. 
	In some cases, developers become inactive after some time or become more focused on a specific project component.
	Therefore, the definition of the \textit{active} developer might be required to get updated from time to time. 
	However, to make our results comparable with those of previous studies, we choose to rely on its common practice and definition. 
	We recommend a rolling approach for the train-test split to overcome outdated decisions for future research. 
	
	Although we study textual information as our independent feature, more external bug characteristics--e.g., the number of comments, keywords, and the number of CC'ed developers--can be added to the prediction models. 
	We plan to expand our independent variables and include additional external factors in future works.
	
	\subsection{Internal validity}
	The bug information is extracted from the Bugzilla using the REST API. 
	We incorporate all the bug records between January 2010 and December 2019 to have an updated bug report information. 
	However, the API is limited for ordinary users, and access to the information of several bugs is not feasible.
	Therefore, we extract all bug comments and complete our dataset using regular expressions. 
	We ensure that our dataset incorporates all publicly available bugs for the {\scshape{EclipseJDT}} project.
	
	\subsection{External Validity} 
	We consider three well-established projects in Bugzilla.
	Although it is compatible with the previous works, our result may not be generalizable to all other open software systems. 
	However, the selected projects are large, long-lived systems, alleviating the likelihood of bias in our report.
	The replication of our study using diverse projects may prove useful. 
	We report different metrics to cover all advantages and disadvantages of the methods. 
	Also, we use SVM as the text classifier following the previous works; however, other classifiers may result in different classification performances.

	\section{Related work}~\label{sec:related_works}
	Several studies have been conducted to automate the bug triage process and decrease the cost of manual bug triage.
	Different techniques have been used to address the problem, e.g., text categorization, tossing graph, fuzzy set-based automatic bug triaging, role analysis-based automatic bug triage, information retrieval, and deep learning techniques.
	
	Automatic bug triage using text classification is proposed by \citet{alenezi2013efficient,anvik2006should,xuan2017automatic} wherein they trained different classifiers such as SVM, Naive Bayes, and C4.5 on the history of bug fixes.
	\citet{xuan2017automatic} enhanced Naive Bayes classifier by utilizing expectation-maximization based on the combination of labeled and unlabeled bug reports. 
	They trained a classifier with a fraction of labeled bug reports. 
	Then, they iteratively classified the unlabeled reports and fitted a new classifier with the labels of all the bug reports. 
	These methods only aim to optimize the accuracy, ignoring many other aspects of the bug triaging process.
	
	\citet{park2011costriage} proposed a method to optimize not only the accuracy but also the cost. 
	This method combines the CBR model with a collaborative filtering recommender (CF) model, enhancing the recommendation quality of either approach alone.
	\citet{alenezi2013efficient} presented a text mining approach to reduce the time and cost of bug triaging. 
	They examined the use of four-term selection methods, namely, log odds ratio, chi-square, term frequency relevance frequency, and mutual information on the accuracy of bug assignment. 
	They aimed to choose the most discriminating terms that describe bug reports. 
	They then built the classifier using the Naive Bayes classifier on bug reports. 
	They also incorporated cost to re-balance the load between developers considering their experience. 
	\citet{kashiwa2020} also emphasized the distribution of the loads among developers. 
	Their method aims to increase the number of bugs fixed by the next release. 
	They formulated the bug triaging process as a multiple knapsack problem that maximizes the developers' preferences given a time limit. 
	Consequently, it mitigates the task concentration to particular, experienced developers.

	\citet{lee2020improving} presented a method that addresses the issues related to LDA fixing time calculation using a multiple LDA-based topic set. 
	Their method improved the existing models by building two additional topic sets, partial topic set (PTS) and feature topic set (FTS). 
	They showed that improved LDA has better classification accuracy. 
	Also, \citet{xia2016improving} recommended a bug triaging method enhanced by specialized topic modeling, named multi-feature topic model (MTM), which extends LDA by considering bugs' components and products. 
	Their proposed approach, TopicMiner, considers the topic distribution of a new bug report to make recommendations based on a developer's affinity to the topics and the feature combination.
	Although these works rely on textual information, they do not consider developer engagement. 
	\citet{ge2020high} proposed a method that overcomes this drawback. They build a high-quality dataset by combining the feature selection and instance selection and studied the impact of developer engagement on bug triage. 
	They considered the product information along with the textual information in the bug report to recommend the best developer for a new bug report.
	
	Many recent studies investigated automating the bug triage using deep learning techniques. 
	\citet{lee2017applying} suggested using a Convolutional Neural Network (CNN) and word embedding to build an automatic bug triage. 
	\citet{mani2019deeptriage} utilized an attention-based deep bi-directional RNN model (DBRNN-A) to automate bug triage. 
	Their approach enables the model to learn the context representation over a long word sequence, as in a bug report. 
	Moreover, they compared their methods with four different classifiers, multinomial naive Bayes, cosine similarity-based classifier, support vector machines, and softmax (regression) based classifier. 
	Their results show DBRNN-A, along with the softmax classifier, outperforms bag-of-words models.
	\citet{guo2020developer} proposed a developer activity-based CNN method for bug triage that recommends a list of developers. 
	They combined CNN with batch normalization and pooling to learn from the word vector representation of bug reports generated by Word2vec.
	
	A significant characteristic of bug reports is their dependency. 
	However, its importance is rarely considered in the bug triage domain. 
	\citet{kumari2019quantitative} developed a bug dependency-based mathematical model to develop software reliability growth models. 
	They interpreted the bug summary description and comments in terms of entropy that also measures the uncertainty and irregularity of the bug tracking system. 
	In the bug triaging process, the incoming bugs are dynamic that makes the bug dependency graph uncertain. 
	To address this issue, \citet{akbarinasaji2018partially} constructed a bug dependency graph considering two graph metrics, i.e., depth and degree. 
	They proposed a Partially Observable Markov Decision Process model for sequential decision making to prioritize incoming bugs based on the bug fixing history and use Partially Observable Monte Carlo Planning to identify the best policies for prioritizing the bugs. 
	
	
	Different from the previous studies, RABT formulates a comprehensive model to capture the most important aspects of bug triage that are specified by the domain experts. 
	RABT uses textual information to estimate the bug fixing time. 
	Also, the information is fed into a classifier, SVM, to find the appropriate developers. 
	However, instead of simply combining these values, RABT considers the importance of developers' available time slots. 
	Given that constraint, it also postpones the bugs that are blocked by others and cannot be solved at the moment. 
	Accordingly, it covers the objectives of the previous works subject to their existing constraints.

	\section{Concluding remarks}~\label{sec:conclusion}
	In this paper, we proposed a dependency-aware bug triaging method that aims to reduce bug fixing time and infeasible assignment of blocked bugs while matching the most appropriate developers. 
	DABT also considers the bug fixing burden of developers in the bug triaging process and alleviates task concentration on a small portion of developers. 
	Accordingly, it reduces the number of overdue bugs before the next release. 
	
	DABT is enhanced by adding the constraints on the blocked bugs. 
	Although it is primarily a triaging method, it also prioritizes the bugs such that both the complexity of the bug dependency graph and the total fixing time reduces.
	Experimenting with three open-source software systems, DABT demonstrated a robust result in terms of the reduced overdue bugs, the improved fixing time of the assigned bugs, and the decreased complexity of the bug dependency graph. 
	The model has lower accuracy compared to the other baselines. However, through sensitivity analysis, we showed that it achieves higher assignment accuracy for different hyperparameter settings.
	
	In this work, we assume that each developer can only work on a single bug simultaneously. 
	It is consistent with previous work; however, this may not be the case in many practical settings. 
	A relevant venue for future research would be to formulate a model that considers multiple knapsacks per developer, each of which has a specific time limitation and working capacity on bugs.

	\section{Supplementary materials}~\label{sec:supplementary}
	To make the work reproducible, we publicly share our originally extracted dataset of one-decade bug reports, scripts, and analysis on \href{https://github.com/HadiJahanshahi/DABT}{\textcolor{blue}{GitHub}}.

	\bibliographystyle{ACM-Reference-Format}
	\bibliography{references}
	

\end{document}